\documentclass[prl,twocolumn,showpacs,floatfix,amsmath,amsfonts]{revtex4}
\usepackage{epsfig}

\begin{document}

\title{Split vortices in optically coupled Bose-Einstein condensates}

\author{Juan J. \surname{Garc\'{\i}a-Ripoll}}
\affiliation{Departamento de Matem\'aticas, Escuela T\'ecnica Superior de
  Ingenieros Industriales, \\
  Universidad de Castilla-La Mancha 13071 Ciudad Real, Spain}
\author{Fernando Sols}
\affiliation{Departamento de F\'{\i}sica Te\'orica de la Materia Condensada,\\
  Universidad Aut\'onoma de Madrid,Cantoblanco, Spain}
\author{V\'{\i}ctor M. \surname{P\'erez-Garc\'{\i}a}}
\affiliation{Departamento de Matem\'aticas, Escuela T\'ecnica Superior de
  Ingenieros Industriales, \\
  Universidad de Castilla-La Mancha 13071 Ciudad Real, Spain}

\date{\today}

\begin{abstract}
  We study a rotating two-component Bose-Einstein condensate in which an
  optically induced Josephson coupling allows for population transfer between
  the two species. In a regime where separation of species is favored, the
  ground state of the rotating system displays domain walls with velocity
  fields normal to them.  Such a configuration looks like a vortex split into
  two halves, with atoms circulating around the vortex and changing their
  internal state as they cross the domain wall.
\end{abstract}


\maketitle


Vortex formation is generally viewed as an unequivocal signature of superfluid
motion in atomic Bose-Einstein condensates. Both in one- and two-component
systems, a number of vortex-like structures have been created and observed
\cite{JILA,vortex-exp} that confirm theoretical predictions
\cite{vortex-theo,freqs,asym}. The Josephson effect between two weakly coupled
condensates is another paradigm of superfluid transport that so far has
received less attention \cite{JILA2,Yale,sols99}. In this article, we present a
combined study of these two fundamental properties of superfluid systems:
vortex and Josephson dynamics. We study the ground state properties of a
rotating double condensate system in which the combined role of vortex
formation and optical coupling between two different hyperfine states gives
rise to a rich physical behavior.

A crucial consequence of the internal Josephson coupling is the
generation of an {\it effective attraction} between both atomic
species due to the energy that atoms gain by choosing a symmetric
mixture of the coupled internal states. Therefore, the most
interesting physics is reached by combining this coupling with
setups that otherwise favor species separation, such as a
particular combination of atomic scattering lengths
\cite{phase-separation,walls} or a separation of the respective
confining potentials \cite{JILA1}. In this type of setups we find
that the effective attraction due to the optical coupling causes
an increase in the thickness of the domain wall where the two
components physically overlap. More important is the fact that, if
we add rotation to a Josephson coupled condensate with separate
domains, atoms can use the domain wall to mutate their internal
state and shift between components in a continous way. Combining
this persistent current in the inner space with a persistent
current in real space, the double condensate may now easily create
a vortex core for each component in the region where that
component has a low density. Even for otherwise small values of
the Josephson coupling, these {\it split vortices} support a net
mass flow comparable to that of conventional vortices. The
formation of these novel structures is less costly in terms of
angular speed because the vortex of a given component is formed
not within its own domain but in the opposite one, where its
superfluid density is low and the cost in kinetic energy is
therefore small.

To study this type of vortex structure, we analyze first a
rotating two-component condensate without optical coupling (i.e.
with impenetrable domain walls) and show that its behavior is
essentially that of a one-component system with a displaced axis
of rotation. Then we show that vortex formation is strongly
inhibited because of the ability of the system to gain angular
momentum by merely distancing itself from the rotation axis. The
picture changes qualitatively when a Raman coupling is introduced
to permit coherent hopping between the two internal states.
Because the flow of particles in a given state is no longer a
conserved quantity, the circulation lines of a component can cross
the domain wall with a concomitant decrease in their supporting
superfluid density. This results in a global structure of two
asymmetrical vortices where matter is efficiently transported both
in real space and within the hyperfine doublet.

\textit{The model.-} In this paper we focus on double condensate systems such
as those made of Rb in JILA \cite{JILA}, but this time in rotating traps and
with a permanent optical coupling between the species. In the rotating frame of
reference \cite{asym} the Gross-Pitaevskii equations for the condensate
wavefunctions read
\begin{subequations}
\begin{eqnarray}
i\hbar\partial_{\tau}\Psi_1 &=&
\left[H_1+{\textstyle \sum_j}U_{1j}|\Psi_j|^2\right]\Psi_1 -
{\textstyle \frac{1}{2}}
\hbar \Omega_R \Psi_2,\\
i\hbar\partial_{\tau}\Psi_2 &=&
\left[H_2+{\textstyle \sum_j}U_{2j}|\Psi_j|^2\right]\Psi_2 -
{\textstyle \frac{1}{2}} \hbar \Omega_R \Psi_1.
\end{eqnarray}
\end{subequations}
Here $H_j$ correspond to the non-interacting Hamiltonians $H_{1,2}
= -\frac{\hbar^2}{2m}\Delta + V({\bf r}-{\bf r}_{1,2}) - \hbar
\Omega L_z + \hbar\bar\delta_{1,2}$, where $V({\bf r}) =
\frac{1}{2}m\omega^2r^2$ is the trapping potential, ${\bf r}_j$ is
the center of the trap for component $j$, and $\Omega$ is the
angular speed. The mutual interactions
$U_{ij}=4\pi\hbar^2a_{ij}/m$, are proportional to the $s$-wave
scattering lengths $a_{ij}$. The last term in each equation models
an optical coupling between species. This is a Josephson-type
coupling that allows atoms to change their internal state. Since
we focus on stationary configurations, the normalization of each
wavefunction is fixed, $N_i = \int |\Psi_i(\mathbf{r})|^2
d\mathbf{r}$. Finally, $\hbar\bar\delta_i$ is a tunable energy
splitting in the hyperfine space, which controls the population of
each component.

\begin{figure}
  \epsfig{width=\linewidth,file=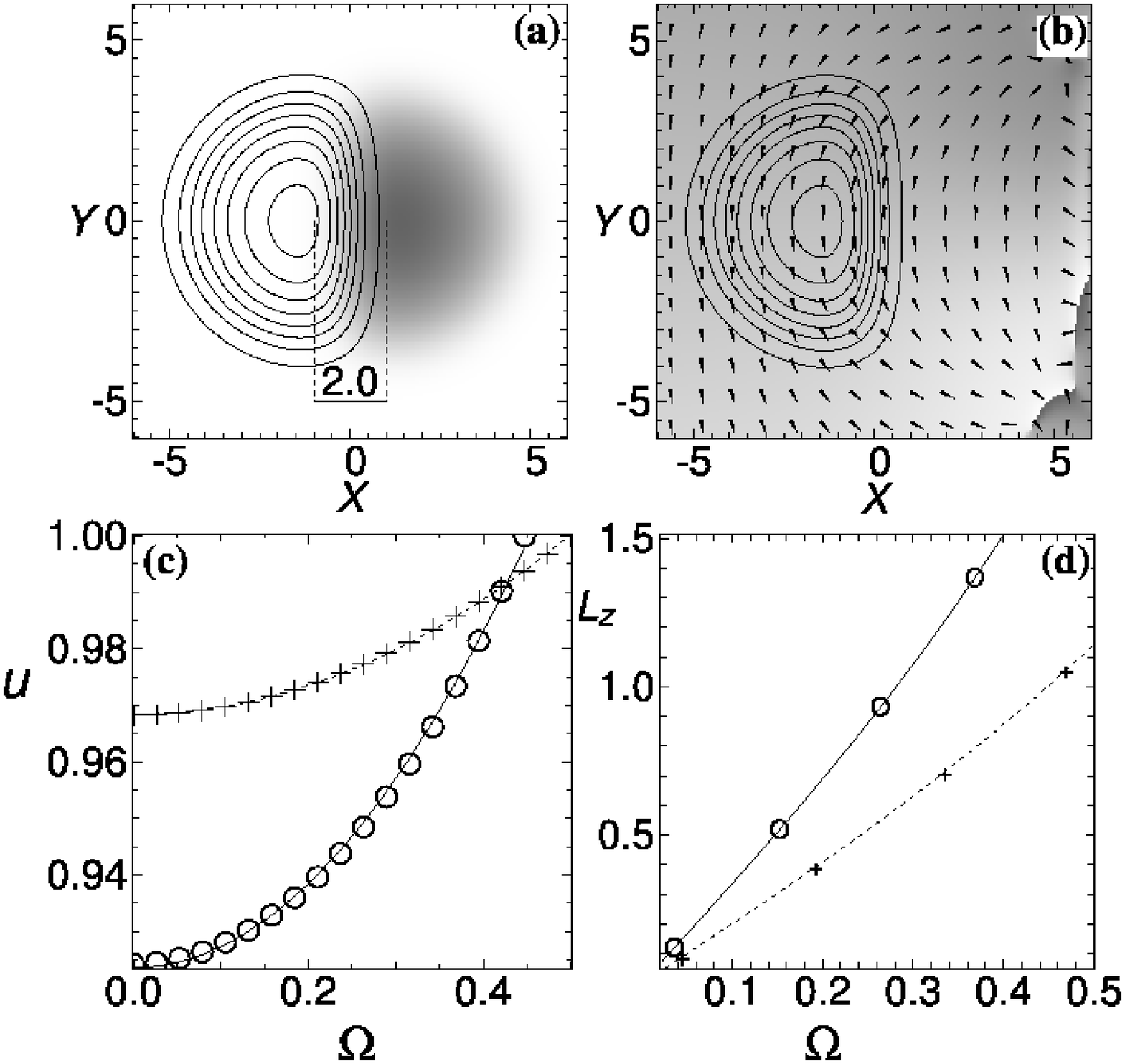}
  \caption{\label{fig-domains}
    (a) Density profiles in scenario A for $\Omega = 0$. We show a grayscale
    plot of $|\psi_1|^2$, (dark), and a contour plot of $|\psi_2|^2$, (solid
    lines).  The trap separation is indicated with dashed lines.  (b) Type A
    condensates for $\Omega=0.135$. Grayscale plot for the phase
    ($\theta=\arg\psi_1$), arrows for the direction of the velocity field
    (${\bf v} =\nabla\theta$), and contour lines for $|\psi_1|^2$.
    (c) Separation of the atomic clouds and (d) angular
    momentum as a function of the angular speed for scenarios A (circles, solid
    line) and B (crosses, dashed line). All magnitudes have been
    adimensionalized.}
\end{figure}

In this paper we restrict our attention to axially symmetric
two-dimensional configurations which may describe pancake type
traps \cite{2D}. In such traps the nonlinear parameter is affected
by a corrective factor \cite{asym,molina} and $N_1$ and $N_2$ are
interpreted as effective number of particles. To ease the analysis
we introduce dimensionless variables $\mathbf{x} = \mathbf{r}/a_0$
and $t = \tau/T$. With this change and $\psi_j(\mathbf{x},t) =
N_j^{-1/2}\Psi_j(\mathbf{r},\tau)$, we write the equations for
stationary states
\begin{subequations}
\label{gpe}
\begin{eqnarray}
 \mu_1\psi_1 &=&
\left[\mathcal{H}_1+\hbox{$\sum_j$}g_{1j}|\psi_j|^2\right]\psi_1 - \lambda \psi_2,\\
\mu_2\psi_2 &=&
\left[\mathcal{H}_2+\hbox{$\sum_j$}g_{2j}|\psi_j|^2\right]\psi_2 -
\lambda \psi_1,
\end{eqnarray}
\end{subequations}
with rescaled Hamiltonians
\begin{equation}
\label{ham2}
  \mathcal{H}_{1,2} = {\textstyle\frac{1}{2}}\left[-\Delta
  +(x - x_{1,2})^2 + y^2\right] + i\Omega
  \left(x\partial_y - y\partial_x\right).
\end{equation}
In Eq. (\ref{gpe}) the splittings $\bar\delta_i$ are included in
the chemical potentials. The parameter $\lambda=2\Omega_R/\omega$
measures the intensity of the optical coupling. Although it is
widely tunable, in our work it will be spatially uniform and at
most of order unity. We have solved numerically Eq. (\ref{gpe}),
looking for the solutions that have lower energy, the so called
ground states. Each of such solutions represent a stable,
experimentally realizable configuration.

We will consider two different scenarios in which the double condensate
exhibits domain walls, and which give qualitatively similar results. The first
case, which we call ``setup B'', corresponds to a situation in which
$g_{11}=g_{22}=N, g_{12}=2N,$ with a choice of $N=38$. In this case the
inequality $g_{12}^2>g_{11}g_{22}$ is satisfied and the domains form
spontaneously \cite{walls} with no need for trap separation nor splitting,
i.~e. $x_{1,2} = 0$, $\mu_1=\mu_2$.

The second and most important scenario, which we call ``setup A" corresponds to
the case of $^{87}$Rb \cite{JILA}
with parameter values $\left(\begin{smallmatrix} g_{11} & g_{12} \\
    g_{21} & g_{22} \end{smallmatrix}\right) = \left(\begin{smallmatrix} 1 &
    0.94 \\ 0.94 & 0.97
\end{smallmatrix}\right) \times \alpha N$ (i.e. $N_1=N_2=N$), and a typical
effective value of $N=100$. This type of condensates has been studied in many
previous experimental and theoretical works. For $^{87}$Rb, the inequality
mentioned above is close to saturation \cite{walls} and a separation
$x_1=-x_2=1$ ensures the formation of two different domains. The external
splitting takes small values, $\mu_1-\mu_2\leq 0.01$, and it does not influence
the results.

\textit{Rotation without Josephson coupling.-} Here we consider
the case $\lambda=0$. The existence of domain walls implies that
the motion of species is spatially constrained. Therefore, if we
impose some angular speed to the traps containing the condensates,
each cloud tends to slip tangentially to the domain wall. The
density distribution is similar to the case without rotation, but
now, due to the centrifugal force, the species separate a little
and gain linear speed. Together with the deformation of the
clouds, this mechanism permits the acquisition of a large amount
of angular momentum without generating vortices [Fig.
\ref{fig-domains}(d)], which only form at very high $\Omega$.

\begin{figure}
{\centering
\epsfig{width=0.40\linewidth,file=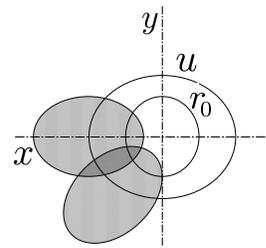}}
\caption{\label{fig:rot-trap}
  Scheme of an off-axis rotating trap. Both the center of the trap and the
  trap itself rotate with angular speed $\Omega$. }
\end{figure}

In Fig. \ref{fig-domains}(a) we show the ground state of the
rotating double condensate system for $\lambda=0$, showing a
structure of domain walls that persists in the presence of
rotation and prevents the formation of vortices, as revealed by
the phase and circulation patterns of Fig. \ref{fig-domains}(b).
From Fig. \ref{fig-domains}(c) it is evident that the both the
mutual repulsion and the rotation of the trap increase the
separation among species. The flow pattern presents a curvature
which is too small [Fig. \ref{fig-domains}(b)], and the gain of
angular momentum [Fig.  \ref{fig-domains}(d)] is due instead to
the displacement of the condensate away from the origin.

To get a deeper understanding of this configuration it is useful to study the
off-axis rotation of a single component condensate.  It experiences a potential
$V({\bf x},t) = \frac{1}{2}({\bf x}-{\bf r}_0(t))A(t)({\bf x}-{\bf r}_0(t))$
with $A(t)$ and $\mathbf{r}_0(t)$ rotating at speed $\Omega$ around a displaced
axis, as depicted in Fig. \ref{fig:rot-trap}. Thanks to a particular symmetry
of the nonlinear Schr\"odinger equation with harmonic potential \cite{us-pre},
any solution can be written as
\begin{equation}
  \label{solution}
  \psi({\bf x},t) = \phi({\bf x}-{\bf u},t) e^{im\dot{\bf u}\cdot {\bf x}/\hbar + f(t)}
\end{equation}
where $\phi({\bf x}-{\bf u})$ is centered on the center of mass
$\mathbf{u}$ and satisfies the Gross-Pitaevskii equation
(\ref{gpe}) for ${\bf r}_0=0$. In the case of radially symmetric
traps, $A=\omega^2$, we find $(1-\Omega^2/\omega^2){\bf u}={\bf
  r}_0$. This tells us that the behavior of a condensate in an off-axis
rotating trap is qualitatively similar to that of a condensate in a centered
trap: Vortices nucleate at similar angular speeds and they all appear inside
the condensed cloud. The difference is that now the condensate has an
additional source of angular momentum due to its displacement with respect to
the origin, $L_z \propto \Omega |u|^2$.

\begin{figure}
  \epsfig{width=\linewidth,file=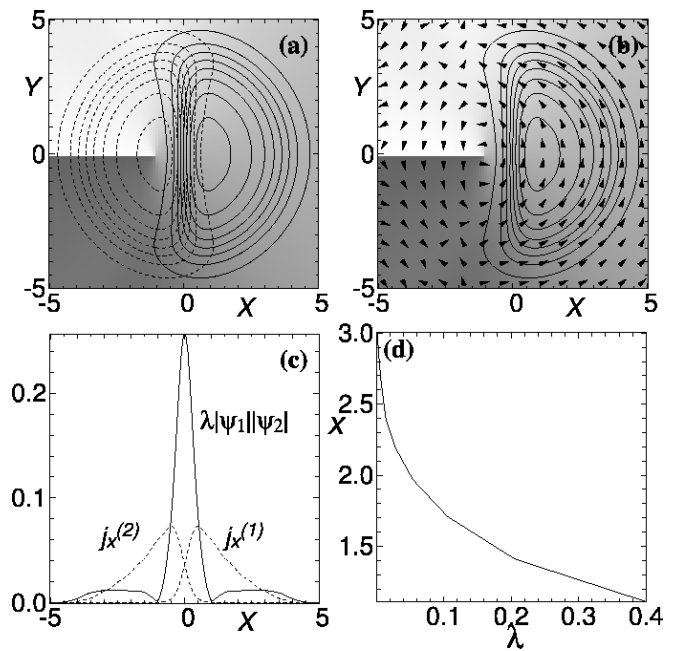}
  \caption{\label{fig-cyclotron}
    Rotating two-component condensate with coupling $\lambda=0.1$,
    scenario B with $N=100$  and angular speed $\Omega=0.2$. (a) Contour
    lines for the moduli of the wavefunctions $|\psi_1|$ (solid line) and
    $|\psi_2|$ (dashed line). (b) Contour levels for $|\psi_1|$, phase
    ($\theta=\arg\psi_1$, grayscale), and velocity field (${\bf
      v}_1=\nabla\arg\psi_1$, triangles).  (c) Condensate fluxes
      $j_x^{(i)}=|\psi_i|^2v_x^{(i)}$, together with
    intensity of the optical coupling, $\lambda|\psi_1|\psi_2|$, along the line
    $y= -0.097$. (d) Position of the vortex for a type B condensate with $\Omega=0.28$,
    as a function of the coupling. All units have been adimensionalized.}
\end{figure}

These arguments, which are rigorous for the single condensate
system, may be extended to the two-component condensate case where
the overlap between clouds is small, meaning that vortices should
appear in the center of the corresponding clouds even when the
trap centers are displaced with respect to the axis of rotation.

\textit{Role of Josephson coupling.-} In order to explain the role
of Josephson coupling let us first consider situation A. It is
easy to show that the Josephson terms favor energetically the
mixing of both species. In the limit of strong coupling, they
coexist in space even when their traps are separated. This is most
intuitively appreciated by inspecting the energy functional
associated to Eq. (\ref{gpe})

\begin{eqnarray}
  \label{energy}
  E[\psi_1,\psi_2] & = & \int \left[ \bar\psi_1 \mathcal{H}_1 \psi_1 + \bar\psi_2 \mathcal{H}_2
  \psi_2\nonumber \right]  \\
  &+& \int \left[  \sum_{i,j =1,2} g_{ij} |\psi_j |^2 |\psi_i |^2
  - \lambda \mathrm{Re}\left(\bar\psi_1 \psi_2\right) \right].
\end{eqnarray}
The coupling term by itself is minimized with a solution such that
$\arg\psi_1=\arg\psi_2$. The analogy with the off-axis rotation of
a single condensate [see Eq. (\ref{solution})] suggests the
variational wavefunction
\begin{subequations}
\begin{eqnarray}
  \psi_1(x,y) &\propto& e^{-(x-u)^2/2-y^2/2} e^{iv y},\\
  \psi_2(x,y) &\propto& e^{-(x+u)^2/2-y^2/2} e^{-iv y}.
\end{eqnarray}
\end{subequations}
Substituting this ansatz into Eq. (\ref{energy}) one gets an effective energy
\begin{equation} \label{energy-variational}
  E \sim {\textstyle\frac{1}{2}}(u-d)^2 + {\textstyle\frac{1}{2}}v^2
  +  \Omega u v - \lambda e^{-u^2} + g_{12}Ce^{-2u^2},
\end{equation}
where $C$ is of order unity. Minimization with respect to the
velocity $v$ leads to $v=-\Omega u$, and thus $\Omega$ favors
separation. More importantly, it is clear from
(\ref{energy-variational}) that $\lambda$ effectively decreases
the repulsion between condensates. Thus, the separation $u$
decreases with the strength of the optical coupling and becomes
zero for a strong enough value of the Josephson coupling,
$\lambda_c$. Both the repulsive interaction among bosons and the
rotation of the trap, tend to inhibit mixture of species, thus
increasing the value of $\lambda_c$.

\begin{figure}
  \epsfig{width=\linewidth,file=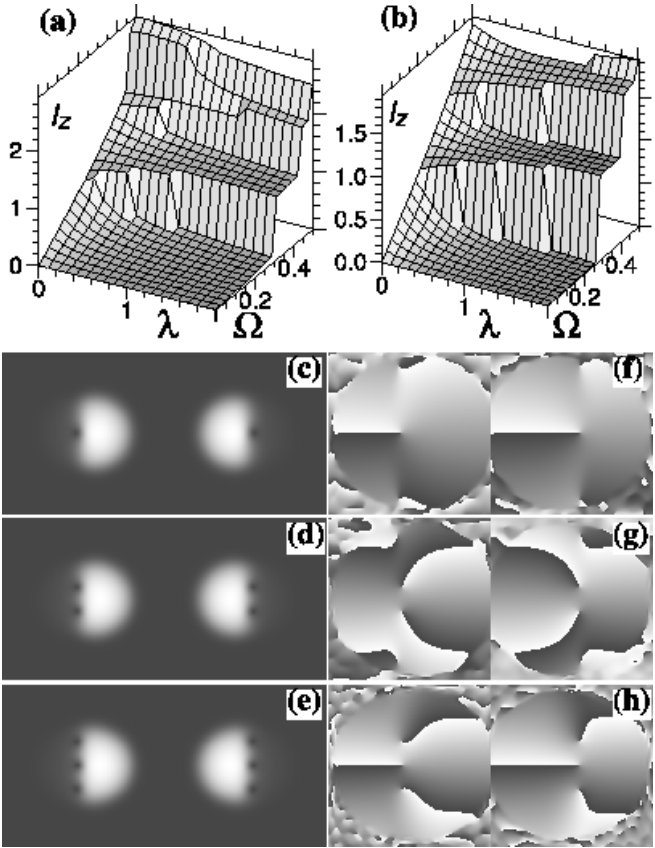}
  \caption{\label{fig-lz-3d}
    (a-b) Angular momentum per particle, $l_z$, versus $\Omega$ and $\lambda$. (a) type A condensates (b) type B condensates. (c-e) Density plots and (f-h) phase plots of
    states of a type A rotating condensate: $\lambda=0.315$, $\Omega=0.316$
    (c,f), $\Omega=0.447$ (d,g) and $\Omega=0.474$ (e,h). In (c-h) we plot
    both species separately (although they overlap). All magnitudes are
    dimensionless.}
\end{figure}

\textit{Vortex formation.-} The structure of matter flow in the
condensate suffers a drastic trasformation when the Josepson
coupling is introduced. From numerical solutions of Eqs.
(\ref{gpe}) for setups A and B, we see that any nonzero coupling
allows the formation of vortices [Figs. \ref{fig-cyclotron}(a) and
\ref{fig-lz-3d}]. These vortices involve a matter flow which is
orthogonal to the wall separating both condensates [Fig.
\ref{fig-cyclotron}(b)]. We note that the exact location of both
vortex cores depends on the intensity of the coupling [Fig.
\ref{fig-cyclotron}(d)].

The mathematical reason for this striking change is that, with a
finite value of $\lambda$, the matter flow of each species is no
longer conserved.  Assuming a divergence free flow, the equation
of continuity along the trajectory of a boson becomes
\begin{equation}
\frac{\partial|\psi_1|^2}{\partial l}v_1 = -\lambda \mathrm{Re}(\bar\psi_1 \psi_2) =
-\frac{\partial|\psi_2|^2}{\partial l}v_2.
\end{equation}
Thus, as the current line is closed around the origin, there is an exchange of
bosons among components. A typical boson flowing around the origin suffers a
transformation from state $|1\rangle$ to $|2\rangle$ and viceversa as it
completes a circle.

Physically, there is a fundamental reason why bosons are
transferred from one component to the other. For $\lambda = 0$,
vortices must lay inside each condensed cloud, due to current
conservation. However, for any nonzero coupling, vortices may
appear {\it outside} the bulk of the clouds, at a variable
distance from the origin. Placed at low density regions, the twist
of the phase requires less kinetic energy
$\int|\psi_i|^2(\nabla\arg\psi_i)^2$ and provides more angular
momentum than the mere separation of clouds.

In Figure \ref{fig-lz-3d}(a-b) we plot the angular momentum per
particle as a function of the Josephson coupling and the angular
speed. The cliffs on the surface are due to the nucleation of
successive vortices.  The separation of these vortices from the
$x=0$ domain wall decreases very fast with increasing $\lambda$:
Fig. \ref{fig-cyclotron}(d) shows that, already for $\lambda
=0.05$, the vortices become visible. As $\lambda\rightarrow 0^+$
for fixed $\Omega$, vortices disappear continuously moving their
core to infinity. To create vortices at $\lambda=0$, one needs to
reach a much higher angular speed, $\Omega_c$, comparable to that
of single component condensates \cite{freqs}. The upshot is that
the optical coupling permits the formation of split vortices with
important mass flow at lower angular velocities.

\textit{Internal Josephson dynamics.-} The Hamiltonian (5) may be
written as that of a nonrigid pendulum \cite{sols99} with an
effective interaction energy $E_c=\sum_{ij} g_{ij} \sigma_{ij}
(-1)^{i+j}$, where $\sigma_{ij}\equiv \int |\psi_i|^2 |\psi_j|^2$,
and an effective Rabi frequency $\omega_R=\Omega_R s_{12}$, with
$s_{12} \equiv \int \bar{\psi_1}\psi_2$. The conclusion is that,
for the setups considered here ($\lambda \alt 0.5$), the internal
two-state dynamics lies in the collective Josephson regime
($2\omega_R/N \ll E_c \ll N\omega_R/2$), while in the JILA
experiment \cite{JILA2}, where no walls are formed, the same
dynamics lies in the noninteracting Rabi limit ($E_c \ll
2\omega_R/N$).

This work has been partially supported by Ministerio de Ciencia y
Tecnolog\'{\i}a under grants BFM2000-0521 and PB96-0080-C02.

\end{document}